\documentclass[conference]{IEEEtran}
\IEEEoverridecommandlockouts

\usepackage{cite}
\usepackage{amsmath,amssymb,bm}
\usepackage{algorithm}
\usepackage{algorithmic}
\usepackage{graphicx}
\usepackage{booktabs}
\usepackage{multirow}
\usepackage{array}
\usepackage{xcolor}
\usepackage{balance}
\usepackage{url}
\usepackage[colorlinks]{hyperref}
\hypersetup{
  citecolor=blue,
  linkcolor=red,
  urlcolor=blue
}

\newcommand{\trans}{\mathrm{T}}
\newcommand{\herm}{\mathrm{H}}
\newcommand{\C}{\mathbb{C}}
\newcommand{\R}{\mathbb{R}}

\newcommand{\junit}{\mathrm{j}}
\newcommand{\norm}[1]{\left\lVert #1\right\rVert}
\newcommand{\abs}[1]{\left\lvert #1\right\rvert}
\newcommand{\Herm}{\operatorname{Herm}}

\title{Sparse Port Selection under Mutual Coupling in Fluid Antenna Arrays}

\author{Jingyuan Xu, Haoyu Liang, Zaichen Zhang, Jian Dang
  \thanks{Jingyuan Xu is with the National Mobile Communications Research Laboratory, Frontiers Science Center for Mobile Information Communication and Security, Southeast University, Nanjing 210096, China. (e-mail: jingyuanxu@seu.edu.cn).}

  \thanks{Haoyu Liang and Zaichen Zhang are with the National Mobile Communications Research Laboratory, Frontiers Science Center for Mobile Information Communication and Security, Southeast University, Nanjing 210096, China. Zaichen Zhang is also with Purple Mountain Laboratories, Nanjing 211111, China. (e-mails: {lianghaoyu, zczhang}@seu.edu.cn).}

	\thanks{Jian Dang is with the National Mobile Communications Research Laboratory, Frontiers Science Center for Mobile Information Communication and Security, Southeast University, Nanjing 210096, China, also with the Key Laboratory of Intelligent Support Technology for Complex Environments, Ministry of Education, Nanjing University of Information Science and Technology, Nanjing 210044, China, and also with Purple Mountain Laboratories, Nanjing 211111, China. (e-mail: dangjian@seu.edu.cn).}

}

\begin{document}

\bstctlcite{IEEEexample:BSTcontrol}

\maketitle
\pagestyle{plain}
\thispagestyle{plain}

\begin{abstract}
Fluid antenna systems obtain spatial degrees of freedom by reconfiguring antenna positions within a confined region, a principle that extends to beamforming: shaped beams can be synthesized using far fewer radio-frequency feeds than candidate antenna positions. When the candidates are densely arranged, however, electromagnetic mutual coupling changes the relationship among terminal voltages, induced currents, and radiated fields, so an uncoupled model no longer describes the hardware and may activate an unsuitable set of ports, distorting the synthesized pattern. This paper develops a mutual-coupling-aware framework that converts the desired beam amplitude into a finite-aperture-compatible complex target and models the complete antenna lattice as a coupled multiport network, selecting the active ports and their source voltages through the coupled voltage-to-field response. Inactive candidate ports remain part of the network and carry induced currents, and every compared design is evaluated through the same electromagnetic model under the same source-voltage budget. Numerical results show that the mutual-coupling-aware design improves both the average mainlobe signal-to-noise ratio (SNR) and the peak sidelobe level (PSLL) over coupling-unaware selection and a fixed array, demonstrating that mutual coupling should be exploited in the design itself rather than compensated only in the final evaluation.
\end{abstract}


\section{Introduction}\label{sec:introduction}

\subsection{Background and Related Work}

Beamforming is central to modern wireless systems: through the
coherent superposition of waves radiated by multiple antenna elements, an
array steers energy toward intended directions while placing nulls
elsewhere, and the mainlobe can approach the full array gain. Existing
designs fall into two complementary categories~\cite{nb1,nb2}. Narrow beams tailored to
individual users offer fine angular resolution, robustness against
interference, and inherent physical-layer security. But their performance
degrades in fast-varying environments and relies on accurate channel state
information. Broad beams underpin synchronization, initial access, and mobility management, where the
channels demand uniform coverage over a macro
sector~\cite{nb2}. An attractive array architecture should therefore be
able to synthesize beams with arbitrary pointing directions and arbitrary
widths within a single framework.

Fluid antenna systems (FASs) provide exactly this flexibility at the
antenna level: the active radiating element is no longer anchored at a
fixed location but can be relocated or switched among candidate positions
inside a finite
aperture~\cite{wong2020limits,new2025tutorial,wong2021fas}. The word
``fluid'' refers to this softwarized adaptability, not liquid
antenna materials, so the effective radiation position becomes
electronically controllable. Most early FAS studies concentrated on the
spatial variation of the channel across the aperture, often referred to as
the random {\em fading gain}~\cite{wt5,zhang2026fbl,wt2,hj3,zzt1}. Because the channel envelope takes a unique value at each candidate
position, moving the active port to a favorable position strengthens the
received signal, enhances reliability, and suppresses interference.

A complementary, broader, and more fundamental dimension of FAS
has recently emerged, the geometric diversity of a fluid
antenna array (FAA), which activates multiple ports
simultaneously~\cite{zhang2026finite,zhang2026planar}. In contrast to
fading-based benefits, this diversity is easy to exploit,
because it comes from a deliberately reconfigurable geometric structure rather
than from the random propagation environment. The reconfigurable geometry
directly determines the effective aperture, the spatial sampling structure,
and the spatial frequency content of the array, and hence its radiation
pattern, so the achievable performance is qualitatively different from
that of fixed apertures. When
candidate positions become densely arranged, mutual
coupling must be included in the model, which has led to
electromagnetic-aware (EM-aware) FAA designs~\cite{zhang2026emaware}.
Designing the port activation of an FAA is thus a geometry-aware
array synthesis problem, not just a channel selection problem.
Driven by this perspective, recent designs include flexible
arbitrary-direction beam synthesis~\cite{11555734}, near-field
beamforming and port selection~\cite{chen2026nearfield}, peak-sidelobe
suppression~\cite{liang2026peaksidelobe}.

When a small set of active ports is to be selected, beam synthesis
becomes a sparse-approximation problem, in which the desired field is
approximated by a few columns of the array-response dictionary, and sparse
recovery algorithms greedily pick the column most correlated with the
current residual and refit the selected coefficients. Such schemes serve
naturally for port and beam selection, including compressive-sensing-based
beam selection in millimeter-wave MIMO~\cite{choi2015beam}, sparse array
synthesis with off-grid refinement~\cite{yang2022sparse}, and discrete
port selection in fluid antenna arrays. They are attractive
whenever the number of radio-frequency feeds is far smaller than the
number of candidate positions, as in classical antenna
selection~\cite{molisch2004}. When the antenna spacing becomes
electrically small, however, the ideal sparse model is incomplete. Mutual
coupling changes the relationship among terminal voltages, antenna
currents, and radiated fields~\cite{gupta1983mutual,balanis2016}, so the
array must be modeled as a coupled multiport network~\cite{zhang2026emaware}.
This matters in a discrete FAA, because selecting the active ports does not remove the remaining candidate antennas. An inactive port has zero source voltage, yet its terminal
current can remain nonzero, induced through the shared impedance network.
The support and the beamforming weights must therefore be designed under
this coupled response, instead of an isolated-element idealization.

\subsection{Challenges and Motivations}

Combining sparse port selection with a densely coupled array model raises two difficulties: the amplitude-only beam specification does not determine a unique complex target for sparse synthesis, and the sparse feeding semantics of the array must be established under mutual coupling. The principal challenges are summarized below.

\begin{enumerate}
\item \textbf{Phase non-uniqueness:} An amplitude-only beam specification does not uniquely determine the complex target field for sparse synthesis. Prescribing the phase in advance reduces the available design freedom and may compromise the sparsity achievable under the same beam mask, while recovering a compatible phase from magnitude-only constraints is itself a nonconvex problem whose solution can depend on initialization~\cite{echeveste2016}. Phase retrieval should therefore be performed before sparse port selection to construct an aperture-compatible complex target~\cite{gerchberg1972}.

\item \textbf{Sparse-voltage electromagnetic semantics:} The sparse variable is the source-voltage vector, whereas the physical current vector is generally dense. Hence, forcing the currents of unselected ports to zero would remove their passive electromagnetic response and would describe a different structure~\cite{gupta1983mutual,balanis2016}. A candidate port should therefore be scored by the coupled field produced per unit source voltage after accounting for coupling, source impedance, and loss~\cite{tropp2007omp,zhang2011robust}, and a coupled reconstruction applied only after ideal support selection cannot recover discarded candidates.
\end{enumerate}

\subsection{Contributions}

This paper combines sparse port selection and a multiport electromagnetic model in one consistent voltage-domain framework. Its main contributions are summarized as follows.

\begin{enumerate}
\item \textbf{Sparse-voltage coupled-array formulation:} We distinguish active ports from candidate antennas and formulate the design in the voltage domain, in which the effective dictionary \(\mathbf A=\mathbf B\mathbf C^{-1}\) maps the source voltages to the far-field samples. Here \(\mathbf B\) is the ideal array-response matrix whose columns are the steering vectors of the candidate ports, and \(\mathbf C\) is the impedance matrix of the complete multiport network, including mutual impedance, source impedance, and loss. Each column of \(\mathbf A\) is therefore the coupled field produced by one unit of source voltage at one candidate port, and the formulation explicitly retains the induced currents on the inactive candidate ports.

\item \textbf{Port-selection framework:} The phase-retrieval stage in this paper is based on the Gerchberg--Saxton algorithm~\cite{gerchberg1972}: it first converts the amplitude-only dual-beam specification into an aperture-compatible complex target. An electromagnetic-aware (EM-aware) OMP then selects the active ports using coupling-aware field atoms, and a common Karush--Kuhn--Tucker (KKT) solver reconstructs the beamforming weights of every compared support under the same source-voltage budget. The complete framework is validated on a representative dual-beam target, and all compared designs are evaluated through the same electromagnetic model.
\end{enumerate}

\emph{Notation:} Bold lowercase and uppercase symbols denote vectors and matrices, respectively. The superscripts \((\cdot)^{\trans}\) and \((\cdot)^{\herm}\) denote transpose and Hermitian transpose. The sets of real and complex numbers are denoted by \(\R\) and \(\C\), and \(\R_+\) denotes the set of nonnegative real numbers. The \(n\)th entry of a vector is \([\mathbf x]_n\), \(\mathcal S\) denotes an index set, and \(\mathbf E_{\mathcal S}\) is the associated column-selection matrix. The Euclidean norm is \(\norm{\cdot}_2\), \(\Herm(\mathbf X)=(\mathbf X+\mathbf X^{\herm})/2\) denotes the Hermitian part of \(\mathbf X\), \(\odot\) denotes the Hadamard product, and \(\mathbf I_N\) is the \(N\times N\) identity matrix.

\section{System Model}\label{sec:system-model}

\subsection{Discrete FAA and Desired Beam}\label{subsec:array-model}

Consider \(N\) physical half-wave dipoles on a planar square candidate lattice
\begin{equation}
\mathcal P=\{\mathbf p_n=[x_n~y_n]^{\trans}\in\R^2\}_{n=1}^{N}.
\end{equation}
Only \(K<N\) candidate ports are active. Their index set is \(\mathcal S\subseteq\{1,\ldots,N\}\), with \(\abs{\mathcal S}=K\). For \(L\) sampled far-field directions \((\theta_\ell,\phi_\ell)\), define
\begin{equation}
u_\ell=\sin\theta_\ell\cos\phi_\ell,~
v_\ell=\sin\theta_\ell\sin\phi_\ell
\end{equation}
and the ideal array-response matrix \(\mathbf B\in\C^{L\times N}\) as
\begin{equation}
[\mathbf B]_{\ell n}
=\exp\!\left(\junit\frac{2\pi}{\lambda}(x_nu_\ell+y_nv_\ell)\right),~
\label{eq:steering}
\end{equation}
If \(\mathbf i\in\C^N\) contains the physical antenna currents, the sampled far-field vector \(\mathbf y\in\C^L\), whose \(\ell\)th entry is the complex far-field amplitude at direction \((\theta_\ell,\phi_\ell)\), is given by
\begin{equation}
\mathbf y=\mathbf B\mathbf i.
\label{eq:field-current}
\end{equation}

Let \(\mathbf g_{\rm am}\in\R_+^L\) specify the desired amplitude pattern, which contains two flat-top beams and a low sidelobe region. Because the desired phase is unspecified, we follow the standard phase-retrieval approach~\cite{gerchberg1972,11555734}: alternating projections generate a phase vector \(\bm\psi\), giving the complex target
\begin{equation}
\mathbf g=\mathbf g_{\rm am}\odot\exp(\junit\bm\psi).
\label{eq:complex-target}
\end{equation}
\subsection{Multiport Electromagnetic Network}\label{subsec:em-network}

Let \(\mathbf Z_{\rm em}\in\C^{N\times N}\) be the reciprocal electromagnetic impedance matrix. Its diagonal entries are the self-impedance \(Z_{\rm self}=73.1+\junit42.5~\Omega\), the classical input impedance of an ideal half-wave dipole~\cite{balanis2016}, and its off-diagonal entries are mutual impedances. For parallel half-wave dipoles separated by \(d_{mn}=\norm{\mathbf p_m-\mathbf p_n}_2\), the induced-electromotive-force approximation gives~\cite{gupta1983mutual,balanis2016}
\begin{align}
Z_{\rm m}(d)
={}&\frac{\eta_0}{4\pi}
\left[2\operatorname{Ci}(u_0)-\operatorname{Ci}(u_+)-\operatorname{Ci}(u_-)\right]
\nonumber\\
&-\junit\frac{\eta_0}{4\pi}
\left[2\operatorname{Si}(u_0)-\operatorname{Si}(u_+)-\operatorname{Si}(u_-)\right],
\label{eq:mutual-impedance}\\
u_0={}&\beta d,~
u_{\pm}=\beta\!\left(\sqrt{d^2+L_{\rm d}^2}\pm L_{\rm d}\right),
\label{eq:mutual-arguments}
\end{align}
where \(L_{\rm d}=\lambda/2\), \(\eta_0\) is the free-space impedance, and \(\operatorname{Ci}(\cdot)\) and \(\operatorname{Si}(\cdot)\) are the cosine and sine integrals. Ohmic loss is represented by \(R_{\rm loss}\mathbf I_N\), so
\begin{equation}
\mathbf Z=\mathbf Z_{\rm em}+R_{\rm loss}\mathbf I_N.
\end{equation}

Every physical port is connected to the same source impedance \(Z_{\rm s}\). The source-voltage vector \(\mathbf v_{\rm s}\) and antenna-current vector \(\mathbf i\) therefore satisfy
\begin{equation}
\mathbf v_{\rm s}
=\underbrace{\left(\mathbf Z+Z_{\rm s}\mathbf I_N\right)}_{\mathbf C}\mathbf i,~
\mathbf i=\mathbf C^{-1}\mathbf v_{\rm s}.
\label{eq:network}
\end{equation}
The key point is that an inactive candidate port obeys \([\mathbf v_{\rm s}]_n=0\), not \([\mathbf i]_n=0\). From \eqref{eq:network}, the zero-source equation for such a port remains coupled to every other current. It can therefore carry an induced current and reradiate.

For completeness, define the radiation, loss, and accepted-power matrices as
\begin{align}
\mathbf R_{\rm rad}&=\Herm(\mathbf Z_{\rm em}),&
\mathbf R_{\rm loss}&=R_{\rm loss}\mathbf I_N,
\label{eq:power-matrices-a}\\
\mathbf R_{\rm acc}&=\mathbf R_{\rm rad}+\mathbf R_{\rm loss}.
\label{eq:power-matrices-b}
\end{align}
Using a consistent root-mean-square phasor convention,
\begin{equation}
P_{\rm rad}=\mathbf i^{\herm}\mathbf R_{\rm rad}\mathbf i,~
P_{\rm loss}=\mathbf i^{\herm}\mathbf R_{\rm loss}\mathbf i,~
P_{\rm acc}=\mathbf i^{\herm}\mathbf R_{\rm acc}\mathbf i.
\label{eq:powers}
\end{equation}

\subsection{Sparse Source-Voltage Formulation}\label{subsec:problem}

For a support \(\mathcal S\), let \(\mathbf q\in\C^K\) collect the nonzero source voltages, which are the complex beamforming weights of the active ports. Then
\begin{equation}
\mathbf v_{\rm s}=\mathbf E_{\mathcal S}\mathbf q,~
\mathbf i=\mathbf C^{-1}\mathbf E_{\mathcal S}\mathbf q.
\label{eq:sparse-voltage}
\end{equation}
Combining \eqref{eq:field-current} and \eqref{eq:sparse-voltage} yields
\begin{equation}
\mathbf y
=\underbrace{\mathbf B\mathbf C^{-1}}_{\mathbf A}
\mathbf E_{\mathcal S}\mathbf q
=\mathbf A_{\mathcal S}\mathbf q,
\label{eq:effective-dictionary}
\end{equation}
where the \(n\)th column of \(\mathbf A\), or atom, is the complete coupled field excited by one unit of source voltage at candidate \(n\). In general, this atom is radiated by all \(N\) physical currents rather than by antenna \(n\) alone.

The joint support and voltage problem is
\begin{equation}
\begin{gathered}
\min_{\mathcal S,\mathbf q}~
\norm{\mathbf g-\mathbf A_{\mathcal S}\mathbf q}_2^2
+\rho\norm{\mathbf q}_2^2\\
\mathrm{s.t.}~
\abs{\mathcal S}=K,~
\mathbf q^{\herm}\mathbf H_{\mathcal S}\mathbf q\leq\Gamma.
\label{eq:joint-problem}
\end{gathered}
\end{equation}
Here \(\rho>0\) is a small ridge parameter. A source-voltage budget uses \(\mathbf H_{\mathcal S}=\mathbf I_K\). An accepted-power budget can instead use
\begin{equation}
\mathbf H_{\mathcal S}
=\mathbf E_{\mathcal S}^{\herm}\mathbf C^{-\herm}
\mathbf R_{\rm acc}\mathbf C^{-1}\mathbf E_{\mathcal S}.
\end{equation}
The cardinality constraint makes \eqref{eq:joint-problem} combinatorial, motivating a greedy support search.

\section{Proposed EM-Aware OMP}\label{sec:proposed-method}

This section first develops the EM-aware OMP port-selection algorithm, in which mutual coupling enters the greedy search through the coupled voltage-to-field dictionary, and then describes the complete end-to-end framework that connects phase retrieval, port selection, and voltage reconstruction.

\subsection{Electromagnetic-Aware OMP Port Selection}

Define the ideal and coupled dictionaries as
\begin{equation}
\mathbf D_0=\mathbf B,~
\mathbf D_{\rm em}=\mathbf B\mathbf C^{-1},
\label{eq:two-dictionaries}
\end{equation}
Coupling-unaware OMP selects its support from normalized columns of \(\mathbf D_0\). This procedure implicitly interprets a coefficient as a directly prescribed current on an isolated candidate and neglects the current redistribution created by \(\mathbf C^{-1}\).

EM-aware OMP instead selects from \(\mathbf D_{\rm em}\). In the considered voltage-limited design, its columns are not normalized: the norm of an atom is physically meaningful because it measures the attainable field per unit source voltage. At iteration \(k\), the candidate port selected is
\begin{equation}
n_k=\underset{n\notin\mathcal S_{k-1}}{\operatorname{arg\,max}}
\abs{\mathbf d_{{\rm em},n}^{\herm}\mathbf r_{k-1}},
\label{eq:omp-selection}
\end{equation}
where \(\mathbf r_{k-1}\) is the current residual. The temporary least-squares estimate and residual are
\begin{align}
\widetilde{\mathbf q}_k
&=\mathbf D_{{\rm em},\mathcal S_k}^{\dagger}\mathbf g,
\label{eq:omp-ls}\\
\mathbf r_k
&=\mathbf g
-\mathbf D_{{\rm em},\mathcal S_k}\widetilde{\mathbf q}_k.
\label{eq:omp-residual}
\end{align}
Because \(\mathbf d_{{\rm em},n}=\mathbf B\mathbf C^{-1}\mathbf e_n\), both the correlation in \eqref{eq:omp-selection} and the residual in \eqref{eq:omp-residual} already contain the induced-current response of every inactive candidate port.

\begin{algorithm}[t]
\caption{Electromagnetic-Aware OMP Port Selection}
\label{alg:em-aware-omp}
\begin{algorithmic}[1]
\REQUIRE Candidate locations \(\mathcal P\), target \(\mathbf g\), active port count \(K\), and network parameters
\ENSURE Support \(\mathcal S_K\) and source voltages \(\mathbf q^\star\)
\STATE Construct \(\mathbf B\), \(\mathbf Z_{\rm em}\), and
\(\mathbf C=\mathbf Z_{\rm em}+R_{\rm loss}\mathbf I_N+Z_{\rm s}\mathbf I_N\)
\STATE Solve \(\mathbf C\mathbf T=\mathbf I_N\) and form
\(\mathbf D_{\rm em}=\mathbf B\mathbf T\)
\STATE \(\mathcal S_0\leftarrow\varnothing\),
\(\mathbf r_0\leftarrow\mathbf g\)
\FOR{\(k=1,\ldots,K\)}
    \STATE Select \(n_k\) using \eqref{eq:omp-selection}
    \STATE \(\mathcal S_k\leftarrow\mathcal S_{k-1}\cup\{n_k\}\)
    \STATE Compute \(\widetilde{\mathbf q}_k\) using \eqref{eq:omp-ls}
    \STATE Update \(\mathbf r_k\) using \eqref{eq:omp-residual}
\ENDFOR
\STATE Reconstruct \(\mathbf q^\star\) on \(\mathcal S_K\) using
\eqref{eq:kkt-solution} and the common source-voltage budget
\RETURN \(\mathcal S_K,\mathbf q^\star\)
\end{algorithmic}
\end{algorithm}

\subsection{Beamforming Framework}\label{subsec:kkt}

\begin{figure*}[t]
\centering
\includegraphics[width=\textwidth]{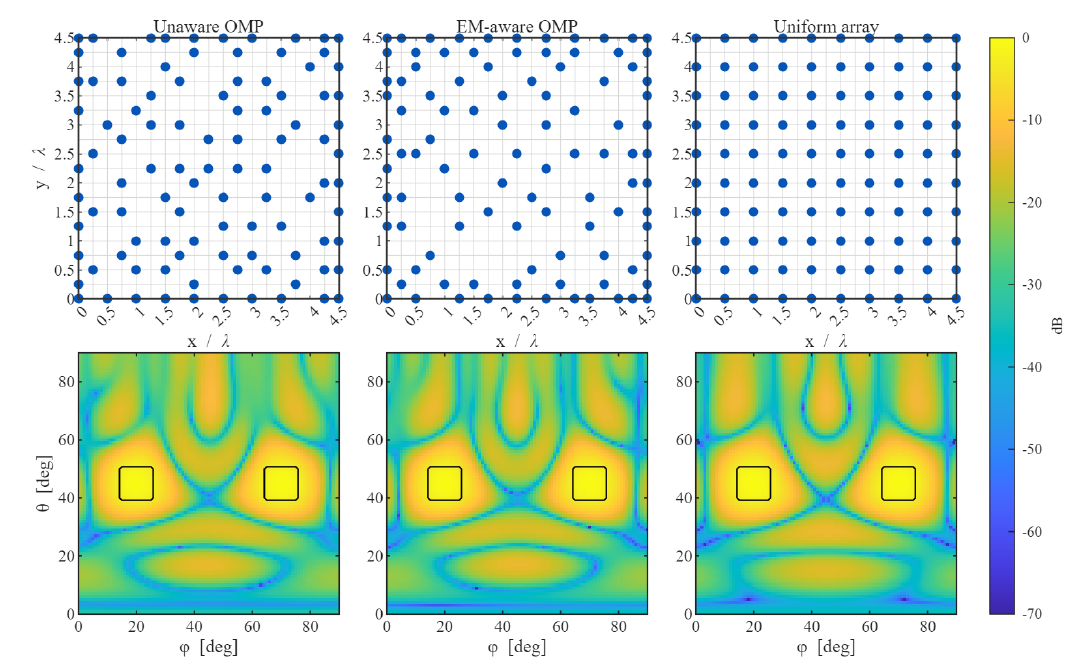}
\caption{Representative active-port layouts and two-dimensional beam patterns. Columns correspond to coupling-unaware OMP, EM-aware OMP, and the uniform array. The top row marks the 100 active ports as solid dots on the \(19\times19\) physical antenna lattice, where unmarked antennas remain terminated members of the coupled network. The bottom row shows the corresponding normalized patterns, and the black outlines indicate the desired dual-beam regions.}
\label{fig:layouts-patterns}
\end{figure*}

The complete framework proceeds as follows. Phase retrieval~\cite{gerchberg1972,11555734} in Section~\ref{subsec:array-model} converts the amplitude mask \(\mathbf g_{\rm am}\) into the aperture-compatible complex target \(\mathbf g=\mathbf g_{\rm am}\odot\exp(\junit\bm\psi)\). Algorithm~\ref{alg:em-aware-omp} then selects the \(K\) active ports on the coupled dictionary \(\mathbf D_{\rm em}\). Finally, the source voltages are reconstructed under the joint problem \eqref{eq:joint-problem}. The temporary OMP coefficients only guide support selection. The final beamforming weights of every compared support are obtained with the same coupled dictionary and budget. For a fixed support, let
\begin{equation}
\mathbf R_{\mathcal S}
=\mathbf D_{{\rm em},\mathcal S}^{\herm}
\mathbf D_{{\rm em},\mathcal S}+\rho\mathbf I_K,~
\mathbf c_{\mathcal S}
=\mathbf D_{{\rm em},\mathcal S}^{\herm}\mathbf g.
\end{equation}
The KKT conditions give
\begin{equation}
\mathbf q(\mu)
=\left(\mathbf R_{\mathcal S}+\mu\mathbf H_{\mathcal S}\right)^{-1}
\mathbf c_{\mathcal S},~\mu\geq0.
\label{eq:kkt-solution}
\end{equation}
If \(\mathbf q(0)\) is feasible, \(\mu=0\). Otherwise, the monotone scalar equation
\begin{equation}
\mathbf q(\mu)^{\herm}\mathbf H_{\mathcal S}\mathbf q(\mu)=\Gamma
\label{eq:dual-equation}
\end{equation}
is solved by bracketing and bisection. The actual currents and evaluation field are finally computed as
\begin{equation}
\mathbf i^\star=\mathbf C^{-1}\mathbf E_{\mathcal S}\mathbf q^\star,~
\mathbf y^\star=\mathbf B\mathbf i^\star.
\end{equation}
This completes the framework from the desired amplitude pattern to the final beam and its evaluation metrics.

\section{Numerical Results}\label{sec:numerical-results}

\begin{figure*}[t]
\begin{minipage}[t]{0.485\textwidth}
\centering
\includegraphics[width=\linewidth]{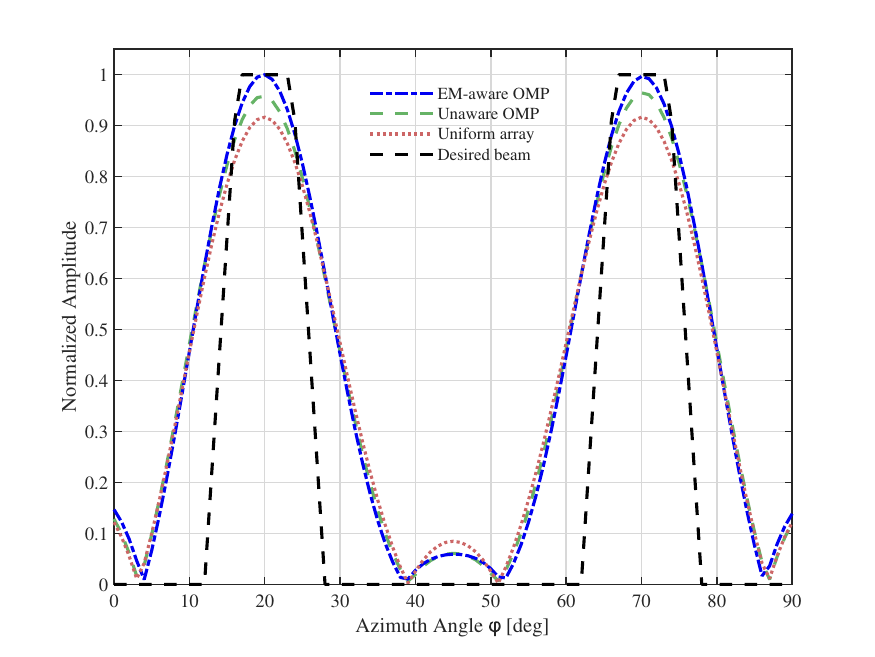}
\caption{Common-normalized azimuth cut at \(\theta=45^\circ\). The shared denominator is the maximum over all three complete two-dimensional fields.}
\label{fig:azimuth-cut}
\end{minipage}
\hfill
\begin{minipage}[t]{0.485\textwidth}
\centering
\includegraphics[width=\linewidth]{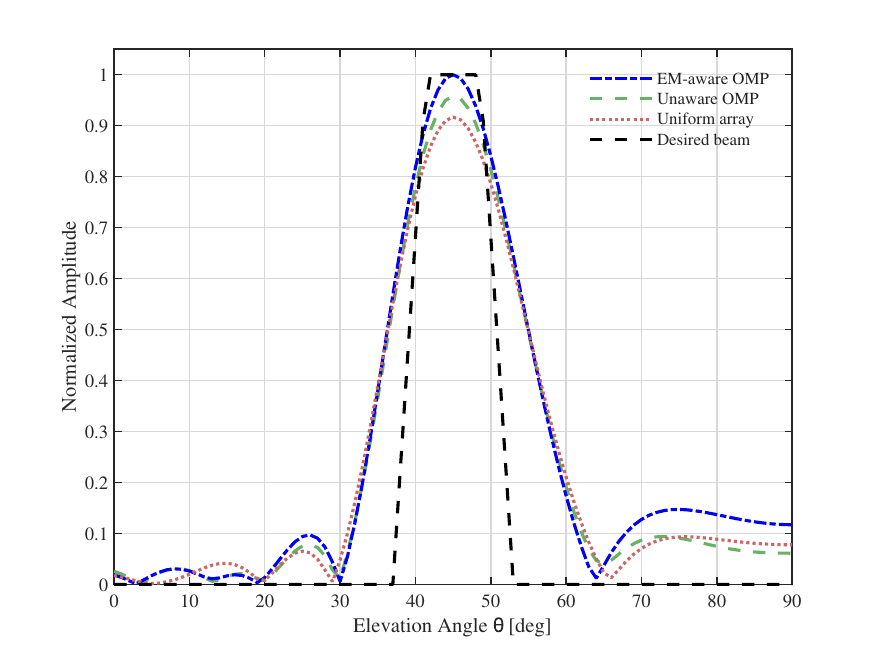}
\caption{Common-normalized elevation cut at \(\phi=20^\circ\), using the same shared two-dimensional denominator as Fig.~\ref{fig:azimuth-cut}.}
\label{fig:elevation-cut}
\end{minipage}
\end{figure*}

\subsection{Simulation Setup}

Table~\ref{tab:parameters} lists the frozen simulation parameters. The square \(4.5\lambda\times4.5\lambda\) aperture contains all \(19\times19=361\) physical half-wave dipoles at \(0.25\lambda\) spacing. Each scheme activates \(K=100\) ports. The uniform baseline occupies every other lattice point and therefore has \(0.5\lambda\) feed spacing. The self-impedance is \(73.1+\junit42.5~\Omega\), and \eqref{eq:mutual-impedance} gives \(Z_{\rm m}=40.79-\junit28.35~\Omega\) at \(d=0.25\lambda\). A minimal \(0.03~\Omega\) diagonal correction makes \(\Herm(\mathbf Z_{\rm em})\) positive semidefinite. The common element factor is omitted to isolate coupling-aware support selection, and \(\norm{\mathbf q}_2\leq40\).

\begin{table}[t]
\caption{Simulation Parameters}
\label{tab:parameters}
\centering
\footnotesize
\renewcommand{\arraystretch}{1.05}
\begin{tabular}{@{}
>{\raggedright\arraybackslash}p{0.36\columnwidth}
>{\centering\arraybackslash}p{0.22\columnwidth}
>{\raggedright\arraybackslash}p{0.32\columnwidth}@{}}
\toprule
Parameter & Symbol & Value\\
\midrule
Wavelength & \(\lambda\) & \(1\)\\
Dipole length & \(L_{\rm d}\) & \(0.5\lambda\)\\
Aperture & \(W_x\times W_y\) & \(4.5\lambda\times4.5\lambda\)\\
Physical candidate lattice & \(N\) & \(19\times19\)\\
Candidate port spacing & \(d_{\mathrm{cand}}\) & \(0.25\lambda\)\\
Number of active ports & \(K\) & \(10\times10\)\\
Self impedance & \(Z_{\rm self}\) & \(73.1+\junit42.5~\Omega\)\\
Source impedance & \(Z_{\rm s}\) & \(50~\Omega\)\\
Loss resistance & \(R_{\rm loss}\) & \(1~\Omega\)\\
Design grid & — & \(61\times61\)\\
Evaluation grid & — & \(91\times91\)\\
Source-voltage budget & \(\norm{\mathbf q}_2\) & \(40\)\\
Noise power & \(\sigma_n^2\) & \(10^{-3}\)\\
\bottomrule
\end{tabular}
\end{table}

The desired pattern contains two flat-top beams with elevation center \(\theta_0=45^\circ\), azimuth centers \(20^\circ\) and \(70^\circ\), half-widths \(\Delta_\theta=\Delta_\varphi=8^\circ\), and flat-top ratio \(0.45\). The three methods are:
\begin{enumerate}
\item \emph{Coupling-unaware OMP}, which selects \(K\) ports from normalized columns of \(\mathbf D_0\).
\item \emph{EM-aware OMP}, which selects from the raw columns of \(\mathbf D_{\rm em}\).
\item \emph{Uniform array}, which uses the fixed \(10\times10\) support.
\end{enumerate}
All methods subsequently use the common reconstruction in Section~\ref{subsec:kkt}. Consequently, the coupling-unaware baseline is unaware only during support selection, while its final field, power, and metrics are still calculated through the same coupled network.

For an evaluation field \(\mathbf y\), the mainlobe and sidelobe sample sets are \(\mathcal M=\{\ell\mid[\mathbf g_{\rm am}]_\ell>0.5\}\) and \(\mathcal L=\{\ell\mid[\mathbf g_{\rm am}]_\ell\leq0.05\}\), respectively. The two reported metrics are the average mainlobe signal-to-noise ratio (SNR) and the peak sidelobe level (PSLL):
\begin{align}
\mathrm{SNR}
&=10\lg\!\left[
\frac{1}{\abs{\mathcal M}}\sum_{\ell\in\mathcal M}
\frac{\abs{y_\ell}^2}{\sigma_n^2}\right],
\label{eq:metric-snr}\\
\mathrm{PSLL}
&=20\lg\!\left(
\frac{\max_{\ell\in\mathcal L}\abs{y_\ell}}
{\max_{\ell}\abs{y_\ell}}
\right).
\label{eq:metric-psll}
\end{align}

\subsection{Result Analysis}

The representative realization has beam centers at
\((\theta,\phi)=(45^\circ,20^\circ)\) and
\((45^\circ,70^\circ)\). The top row of Fig.~\ref{fig:layouts-patterns} shows the active port locations. Every grid intersection is still a physical antenna. Solid dots identify the active ports. EM-aware OMP selects a visibly different support because each candidate is evaluated through the global network response rather than through an isolated steering vector.

The bottom row of Fig.~\ref{fig:layouts-patterns} shows that every method forms both requested beams, while EM-aware OMP provides stronger useful field and better sidelobe control. For this realization, the average mainlobe SNRs of coupling-unaware OMP, EM-aware OMP, and the uniform array are \(22.93\), \(23.21\), and \(22.60\) dB, respectively. Their PSLLs are \(-3.71\), \(-3.83\), and \(-3.39\) dB.

Figs.~\ref{fig:azimuth-cut} and~\ref{fig:elevation-cut} present the \(\theta=45^\circ\) azimuth cut and the \(\phi=20^\circ\) elevation cut, respectively. To preserve absolute differences, all curves are divided by one common denominator: the largest amplitude attained by any method over its complete two-dimensional evaluation surface. The denominator is therefore not recomputed on either displayed cut or separately for each method. EM-aware OMP is consequently seen above the two baselines around both desired beams in the azimuth cut and around the first beam in the elevation cut, whereas independent peak normalization would have hidden most of this gain.

EM-aware OMP improves the average mainlobe SNR by \(0.28\) dB over coupling-unaware OMP and by \(0.61\) dB over the uniform array, while improving the PSLL by \(0.12\) dB and \(0.44\) dB, respectively. The gains are consistent across both metrics, and the SNR gain over the uniform array is the most substantial one. Finally, when the off-diagonal coupling scale is set to zero, \(\mathbf C^{-1}\) reduces to a common diagonal scaling and the aware and unaware OMP supports produce identical results, confirming that the measured gain originates from integrating mutual coupling into support selection.

\section{Conclusion}\label{sec:conclusion}

This paper developed a mutual-coupling-aware framework for discrete FAA beamforming. The formulation treats the port weights as source voltages, retains all physical antennas in the coupled network, and therefore allows the inactive candidate ports to carry induced currents. Mutual coupling enters the greedy search through the effective dictionary \(\mathbf B\mathbf C^{-1}\), rather than being appended only during final evaluation. A common KKT voltage reconstruction then isolates the effect of support selection. Numerical results on a representative dual-beam target show consistent improvements in average mainlobe SNR over coupling-unaware OMP, together with larger gains over a uniform array. These results demonstrate that coupling can be exploited as a structured part of the beamforming dictionary. Future work can incorporate measured or full-wave impedance matrices, broadband matching networks, robust impedance uncertainty, and joint selection under multiple simultaneous hardware budgets.

\bibliographystyle{IEEEtran}
\bibliography{references}

\end{document}